\documentclass[twocolumn]{aa}
\usepackage{graphicx}
\usepackage{txfonts}
\begin{document}
   \title{Planetary nebulae with emission-line central stars}

   \author{
           K. Gesicki, \inst{1}
           A. A. Zijlstra, \inst{2}
           A. Acker, \inst{3}
           S. K. G\'orny, \inst{4}
           K. Gozdziewski, \inst{1}
          \and
           J. R. Walsh \inst{5}
          }

   \offprints{K. Gesicki}

   \institute{
           Centrum Astronomii UMK, 
           ul.\,Gagarina 11, PL-87-100\, Torun, Poland \\
           \email{Krzysztof.Gesicki@astri.uni.torun.pl, 
              Krzysztof.Gozdziewski@astri.uni.torun.pl}
         \and
           School of Physics and Astronomy, University of Manchester,
           PO Box 88, Manchester M60 1QD, United Kingdom \\
           \email{a.zijlstra@umist.ac.uk}
         \and
       Observatoire astronomique de Strasbourg, 11 rue de l'Universite,
           67000 Strasbourg, France\\
           \email{acker@astro.u-strasbg.fr}
         \and
            Copernicus Astronomical Center, ul.Rabianska 8, 
            PL-87-100 Torun, Poland\\
            \email{skg@ncac.torun.pl}
         \and
           Space Telescope European Co-ordinating Facility, ESO,
           Karl-Schwarzschild-Strasse 2, 85748 Garching, Germany\\
           \email{jwalsh@eso.org}
             }

   \date{Received ; accepted }

   \abstract{The kinematic structure of a sample of planetary nebulae,
     consisting of 23 [WR] central stars, 21 weak emission line stars ({\it
       wels}) and 57 non-emission line central stars, is studied.  The [WR]
     stars are shown to be surrounded by turbulent nebulae, a characteristic
     shared by some {\it wels} but almost completely absent from the
     non-emission line stars.  The fraction of objects showing turbulence for
     non-emission-line stars, {\it wels} and [WR] stars is 7\%, 24\%\ and
     91\%, respectively. The [WR] stars show a distinct IRAS 12-micron excess,
     indicative of small dust grains, which is not found for {\it wels}. The
     [WR]-star nebulae are on average more centrally condensed than those of
     other stars.  On the age-temperature diagram, the {\it wels} are located
     on tracks of both high and low stellar mass, while [WR] stars trace a
     narrow range of intermediate masses. Emission-line stars are not found on
     the cooling track. One group of {\it wels} may form a sequence {\it
       wels}--[WO] stars with increasing temperature.  For the other groups
     both the {\it wels} and the [WR] stars appear to represent several,
     independent evolutionary tracks. We find a discontinuity in the [WR]
     stellar temperature distribution and suggest different evolutionary
     sequences above and below the temperature gap. One group of cool [WR]
     stars has no counterpart among any other group of PNe and may represent
     binary evolution. A prime factor distinguishing {\it wels} and [WR]
     stars appears to be stellar luminosity. We find no evidence for an
     increase of nebular expansion velocity with time.

 }

   \authorrunning{K.Gesicki et al.}
   \titlerunning{Planetary nebulae with emission-line central stars}
   \maketitle
%
%________________________________________________________________

\section{Introduction}

Some central stars (cores) of planetary nebulae (PNe) show broad, stellar
emission lines similar to Wolf-Rayet stars. Although superficially similar,
they differ from classical WR stars in their degenerate structure, much lower
masses, a wider range of temperatures, and being limited almost exclusively to
carbon-rich stars (there are very few counterparts to the massive WN stars).
Similar to their massive WC-star counterparts, they are deficient in hydrogen.
They form a class named [WC]-type which historically was further divided into
early [WCE] and late [WCL] groups. A less numerous class of PNe central stars
show narrower and weaker emission lines than the [WR]-type stars. These are
named {\it wels} (weak emission-line stars), as defined in Tylenda et al.
(1993).

The evolutionary status of the [WR]-type stars is still very uncertain, and it
is unclear whether there is any evolutionary relation to the {\it wels}.
Nebular data suggest an evolutionary sequence [WC11]$\rightarrow$[WO2]
(Zijlstra et al.  1994, Acker et al. 1996, Pe\~na et al. 2001), followed by
PG1159 stars (Werner et al. 1992).  Acker \& Neiner (2003) propose a sequence:
[WC11]$\rightarrow$[WC4]$\rightarrow$[WO4]$\rightarrow$[WO1].  Parthasarathy
et al. (1998) have suggested that the {\it wels} are related to the PG1159
stars: [WCL]$\rightarrow$[WCE]$\rightarrow${\it
  wels}/PG1159$\rightarrow$PG1159.  But this sequence is by no means proved;
the precise location of the {\it wels} in relation to the [WR] stars is under
discussion (Marcolino \& de Araujo 2003). Pe\~na et al.
(2001) also argue against too closely identifying {\it wels} with PG1159 stars
(see also Koesterke 2001).  Not all PG1159 are hydrogen-poor (Dreizler et al.
1996), showing that not all PG1159 stars will have evolved from [WR] stars.
Neither are all {\it wels} hydrogen-poor: a dual evolutionary sequence may
be expected.

The lack of hydrogen is often taken to indicate that the [WR] stars have
undergone a late thermal pulse (or helium flash), either during the post-AGB
evolution (a so-called Late Thermal Pulse or LTP) or on the white dwarf
cooling track (Very Late Thermal Pulse or VLTP: Herwig 2001, Hajduk et al.
2005). Following such a pulse, the star rejuvenates and retraces part of its
earlier evolution.  This predicts that nebulae around [WR] stars should be
more evolved than around non-[WR] stars, but this is not confirmed by
observations; the properties of the PNe around [WR] stars do not differ from
those around other central stars (G\'orny 2001).

In this paper we discuss a much larger sample of [WR] stars and {\it wels} than
has previously been available. We apply velocity-field analysis to locate the
objects on derivatives of the HR diagram. In Sect.\,2 we describe the methods
applied for analysis of PNe with specific attention towards automatic model
fitting. Sect.\,3 presents the 33 newly analyzed PNe. In Sect.\,4 we discuss the
nebular and stellar parameters for the full sample of 101 PNe. In Sect.\,5 we
discuss possible evolutionary relations.

Crowther et al. (1998) have refined the WC and the WO  schemes
used for WR stars and define a unified classification for  massive WR
and low-mass [WR] stars. Acker \& Neiner (2003) developed this  scheme
for a sample of 42 PN central stars, classified into [WO 1-4] and  [WC
4-11] stars. This classification (used in the present paper) is based on
the ionization level of the  elements (depending on the wind-temperature
of the PN core), showing essentially carbon lines for the coolest stars
and oxygen lines for the hottest ones.

\section{Methods}

We derive information on the nebulae using a combination of line
ratios, diameters, and high resolution spectra. The diameters and line
ratios are used to fit a photo-ionization model. The model is
constrained to reproduce the intensity distribution of images, if any
are available. Line profiles are obtained from the high resolution
spectra, and are fitted using the emissivity distributions of the
photoionization model, and assuming a velocity field. This is done
using the Torun models (Gesicki et al. 1996, 2003).  The velocity fields
include separate contributions from expansion and turbulence.
Turbulence always includes the instrumental broadening and the thermal
broadening (calculated from the photoionization model), but some
objects may require additional turbulence.

The models assume spherical symmetry. Some strongly bipolar objects
cannot be fitted, because they show very irregular velocity profiles.
But the majority of objects can be well reproduced with a spherical
model. The model corrects for the size of the aperture used for the
spectroscopy, and includes seeing effects. 

The diameters and distances are adopted from the literature. The models
find the density distribution, stellar temperature and the velocity field. The
density as function of radius is usually assumed in the shape of a reverted
parabola, however if  images were available we compared the computed surface
brightness and improved the run of the density. The stellar temperature
assumes a black-body spectrum energy distribution -- therefore we prefer to call
it $T_{\rm b-b}$ instead of $T_{\rm eff}$. This assumption can be
challenged, especially for the [WR] central stars where the comparison between
the  stellar temperature $T_*$ obtained from non-LTE modelling and $T_{2/3}$
which refers to the radius $R(\tau_{\rm Ross} = 2/3)$ shows differences. The
differences are significant for hotter [WO] stars (Koesterke \& Hamann 1997)
while smaller for cooler [WC] objects (Leuenhagen et al. 1996). The general
tendency is always the same: the $T_{2/3}$, $T_{\rm eff}$ or Zanstra
temperatures are smaller than $T_*$. Despite this discrepancy we apply the
$T_{\rm b-b}$ for further discussion since it has the advantage of providing a
uniform measure of the temperature over a range of stellar classes.

The velocity distribution is assumed to vary arbitrarily and smoothly with
radius. In Gesicki \& Zijlstra (2000) we compared the Torun model analysis
with more traditional methods of deriving the expansion velocities from the
spectra. Since long time it was known that different observed lines resulted in
different velocities (e.g. Weinberger 1989) indicating a velocity gradient. Our
model has this advantage that it combines these data into a single velocity
curve. Earlier it was not obvious but the Torun models have shown that even the
shape of a single line can indicate a velocity gradient. Good velocity fields
can be obtained if a larger number of lines are available. The resulting fields
show detailed structure, with velocity peaks at the outer and sometimes at the
inner radius of the nebula (Gesicki \&\  Zijlstra 2003). If fewer lines are
available, the velocity field is less constrained and simplifying assumptions
need to be made. From the full velocity curve $V(r)$ we derive a single
parameter characterizing the nebular expansion. We define the expansion
velocity as a mass-weighted average over the nebula, $V_{\rm av}$. This
parameter has been shown to be robust against the simplifying assumptions: it
can be accurately determined even when the velocity field itself is uncertain
(Gesicki et al 2003). This allows us to define a kinematic age to the nebula.

\subsection{The genetic algorithm}

\begin{figure}[ht]
   \includegraphics[width=88mm]{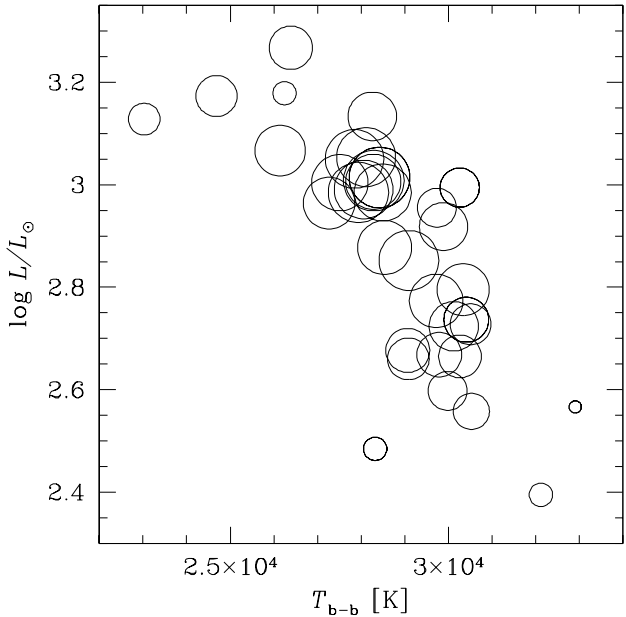}   
      \caption{The best fit luminosity and temperature for photoionization 
     models for a single object (He\,2-113, PN\,G\,321.0+03.9)  for 40 different
     PIKAIA runs. The larger the size of the circle the better the fit quality.}
         \label{ltfits}         
\end{figure}

The Torun models were recently adopted to automate the search for a best-fit
model, using an optimization routine. We applied the method widely known as
the genetic algorithm. While it is still not a very popular optimization
technique, it has proved to be very effective and robust in many problems, in
particular of astrophysical origin. For a review, we refer to a paper by
Charbonneau (1995) who is the author of the publicly available code PIKAIA for
a genetic
algorithm\footnote{http://www.hao.ucar.edu/Public/models/pikaia/pikaia.html},
used for computations in this work.

The genetic algorithms have been invented as an optimization technique
mimicking the processes of biological evolution (e.g. Koza 1992).  They lead
to the selection and adaptation of life forms to the conditions of the natural
environment.  In this sense the evolution can be thought of as a powerful
optimization algorithm. The genetic optimization starts with a set
(population) of randomly chosen individuals (parameters of the mathematical
model encoded in the `genomes'). The whole parameter space of the problem
creates the environment for these individuals. The likelihood of survival of a
given individual is determined by its `fitness' function $f$; usually, in the
least squares minimization, $f=1/\sqrt{\chi^2}$. Then the fitness of a
particular individual is a measure of goodness of fit to the studied data set.
The population evolves through a number of generations. At every generation,
the genetic algorithm evaluates $f$ resulting from each parameter set and
processes information by applying to the genomes genetic operators like
crossover, mutation and selection.  The best fitted members of the population
are used to produce a new generation and the process continues until some
convergence criteria is reached. Thanks to keeping the memory of the best
fitted individuals, the genetic algorithms are much more efficient than Monte
Carlo based search techniques. They are robust and especially well suited for
multi-dimensional problems, models possessing multiple local extrema and
discontinuities. They are non-gradient methods and thus especially well suited
for models dependent in a sophisticated way on their parameters.  Details of
practical application of the technique are given in an excellent introduction
by Charbonneau (2002).

The genetic optimization replaced the 'trial and error' search of parameters
applied in our earlier publications (e.g. Gesicki \& Zijlstra 2003, Gesicki et
al. 2003). The algorithm appeared very effective in comparison with the earlier
work. We note also that because of random initialization of the population
the method is now free from initial biases. The search procedure again follows
two steps. First PIKAIA searches for the values of the nebular mass and the
stellar temperature and luminosity which optimize the fit to the observables,
before attempting to find the velocity field which optimizes the fit to the line
profiles. The following expression is optimized: 

\[
\chi^2 = 
\frac{1}{N-p-1}\sum_{i=1}^N \left[ \frac{O(i) - C(i)}{\sigma(i)}\right]^2, 
\] 

\noindent where $O(i)$ are $N$ observations ($i=1,\ldots,N$),  $C(i)$ are model
predictions dependent on $p$ parameters and $\sigma(i)$ are individual errors
of the observations. In some cases considered in this work, $N$ is very small
and the determination of uncertainties on the fitted is difficult (and without
much meaning in terms of the formal, statistical approach).  Therefore we
did not perform a formal error analysis of the results.
Nevertheless, in a well performed optimization search the
error analysis is as important as the determination of the fit parameters.

To check the possible degeneracy of the solutions and to obtain an idea
about the errors, we collected the values to which the search process converged
in many independent runs with the same start, and projected them on
chosen parameter planes.  The distribution of these parameters gives insight on
the significance of the minima, whether they can be well localized and fixed in
the parameter space of the problem.  In Fig.\,\ref{ltfits} we present an example
of fitting the photoionization model for a single PN (He\,2-113,
PN\,G\,321.0+03.9).  Results are plotted for 40 different runs of PIKAIA, using
the same set of observables.  A scatter in the obtained parameters is seen: few
of the points are overlapping.  The size of the symbols corresponds to the fit
quality.  For this PN the temperature is better constrained than the stellar
luminosity.  The situation for other PNe is similar. For higher temperatures the
models seem to be better constrained. We estimated the accuracy in $T_{\rm b-b}$
as $\pm 2000 \rm K$ and in $\log L/L_{\odot}$ as $\pm 0.3$.

Similar test were performed for the velocity procedure. When fitting a
simple velocity field (linear or parabola-like) the PIKAIA routine
repeats very well the results. More complicated velocity fields result in
different shapes for different runs; however the largest discrepancies appear in
the least constrained parts of the velocity curves. Nevertheless the
mass-averaged expansion velocity remains well determined and is similar in all
runs.  From the distribution of models, we estimate the accuracy of $V_{\rm av}$
as about $\pm 2{\rm km\,s^{-1}}$.

\subsection{Observations}

  \begin{figure}[ht]
   \includegraphics[width=88mm]{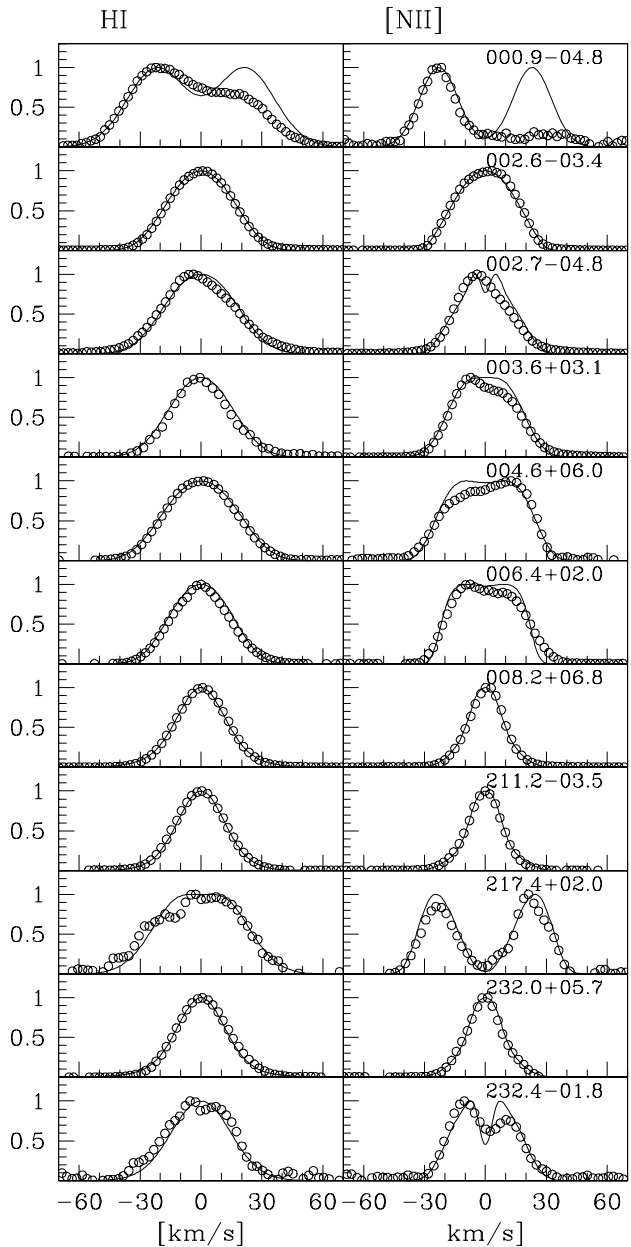}   
      \caption{The observed and modelled lines $\ion{H}{i}$ 6563\AA\ and 
        [$\ion{N}{ii}$] 6583\AA.  The circles correspond to the observed
        profile, the line to the fitted model. The line fluxes are normalized to
        unity. The X-axis velocity scale given in the lowest boxes is the same
        for all plots.  }
         \label{li_1}
   \end{figure}

   \begin{figure}[ht]
   \includegraphics[width=88mm]{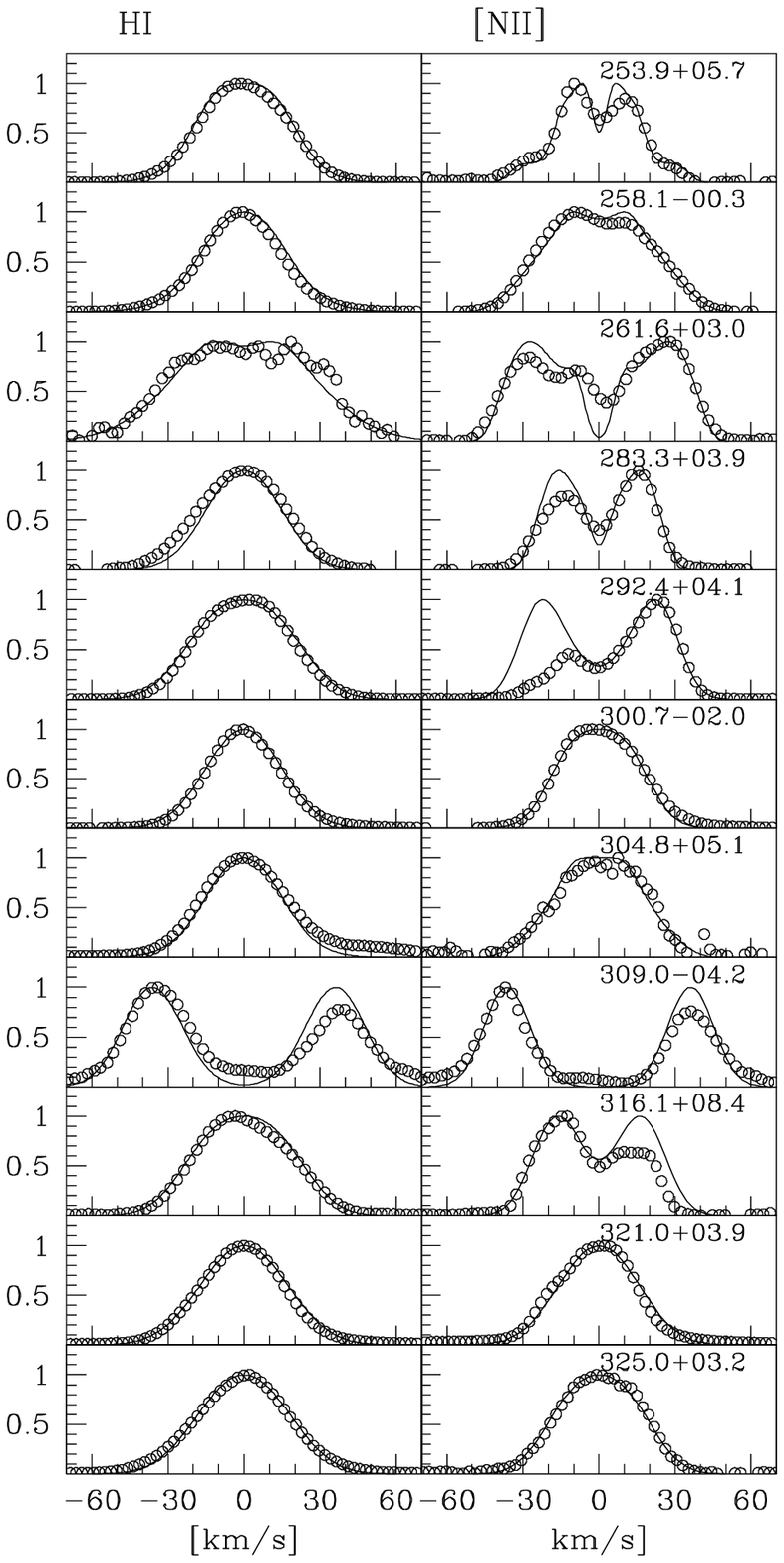}
      \caption{The observed and modelled lines $\ion{H}{i}$ and 
        [$\ion{N}{ii}$].  Continuation of Fig.\,{\ref{li_1}}}
         \label{li_2}
   \end{figure}

   \begin{figure}[ht]
   \includegraphics[width=88mm]{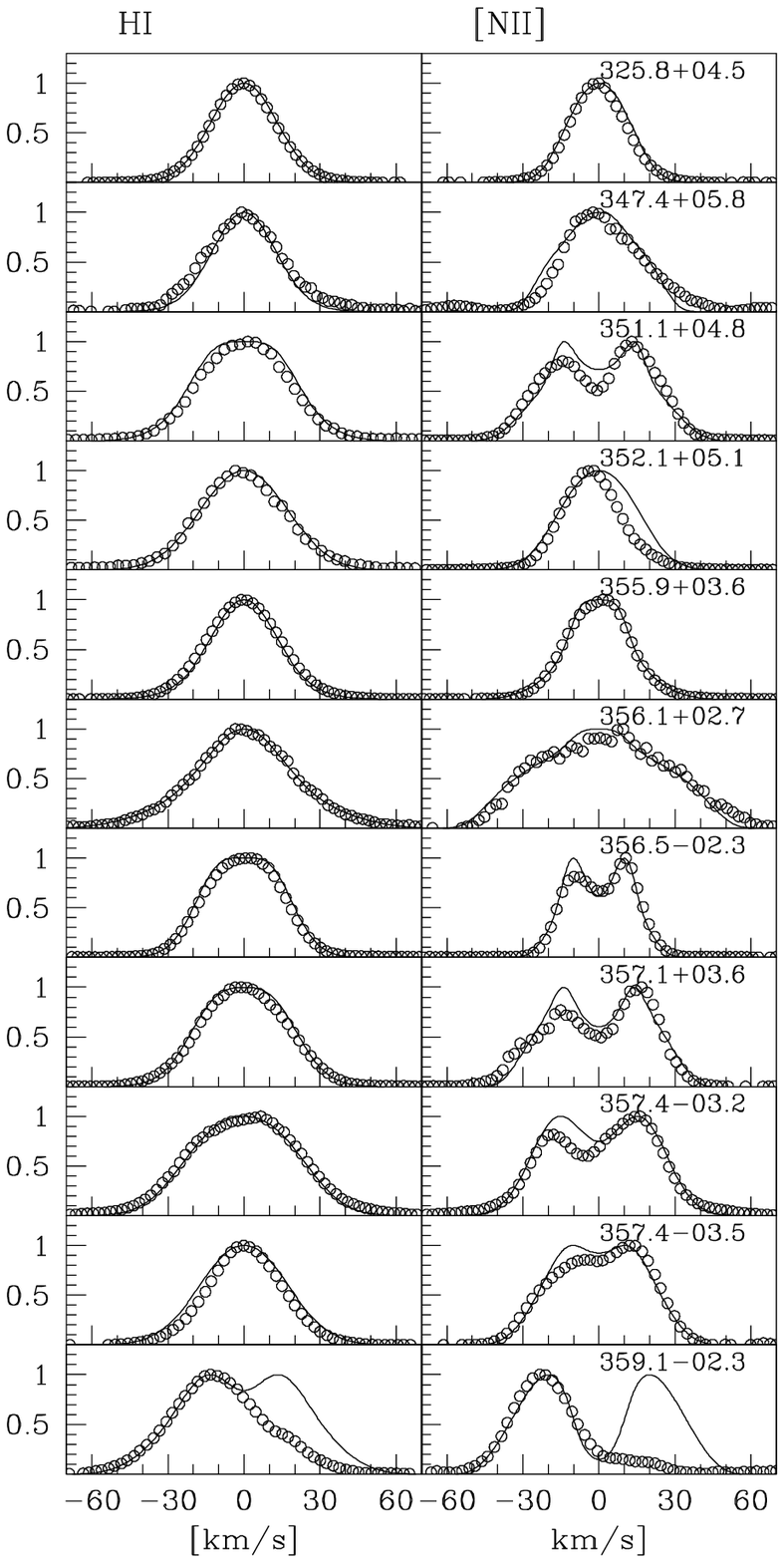}
      \caption{The observed and modelled lines $\ion{H}{i}$ and 
        [$\ion{N}{ii}$].  Continuation of Fig.\,{\ref{li_2}}}
         \label{li_3}
   \end{figure}
   
A large number of PNe have been analyzed in previous papers using the Torun
models. In most cases the velocity fields were determined from three lines
(hydrogen H$\alpha$ 6563\AA, [\ion{N}{ii}] 6583\AA\ and [\ion{O}{iii}]
5007\AA) or fewer.  To this sample we add in the present paper 27 unpublished
CAT observations, which cover only H$\alpha$ and [\ion{N}{ii}] 6583\AA,
supplemented with recent echelle NTT observations of 6 PNe with emission-line
central stars. PNe with new data are listed in Table 1.

The ESO Coud\'e Auxiliary Telescope (CAT) was a subsidiary 1.4m telescope
feeding into the Coud\'e Echelle Spectrograph (CES) located at the
neighboring 3.6m telescope.  The CAT observations of 27 objects were performed
during 1993 and 1994.  The long camera was used, giving a spectral resolution of
60\,000 (corresponding to 5\,km\,s$^{-1}$). The slit width was 2\arcsec.  The
observations use a long slit, sampled with 2\arcsec\ pixels.  However, the
spectra analyzed here use only the central row of pixels.  (As the nebulae
studied here are compact and the CAT was not an imaging-quality telescope, no
spatial information was expected.)  The spectrum covers one order of the
echelle, covering H$\alpha$ and the [\ion{N}{ii}] lines at 6548\AA\ and 6853\AA\
but no other nebular lines.

Six objects with emission-line central stars were observed with the ESO New
Technology Telescope (NTT) during June 2001. The echelle spectra were obtained
using ESO Multi Mode Instrument (EMMI), with grating 14 and cross
disperser 3, giving a resolution of 60\,000. The slit width was 1\arcsec. The
spectra covered the wavelength range 4300--8100\AA. The spectra were summed over
the slit length of 3\arcsec. Exposure times were typically 120 seconds. Between
3 and 7 lines per object showed high enough S/N to be used in the velocity
analysis. The details of the data reduction are given in Gesicki \& Zijlstra
(2003).

The profile fitting procedure requires specification of the center of the
emission line with zero velocity. We did not perform the full radial velocity
correction. Instead we assumed the zero position in the centre of symmetry for
the almost symmetric lines and usually located the zero position in the centre
between the two maxima of the asymmetric lines. In the second case the presence
of two (or more) spectral lines helped to locate the zero position, the
obviously hopeless cases were not analyzed.

\section{The modelled planetary nebulae}

\begin{table*}
\caption[]{The expansion velocities and other data concerning the 
nebulae and the central stars, based on observations described in this paper. 
The column 'remarks' indicate for
PNe observed with EMMI, how many emission lines were used for the analysis.}
\begin{flushleft}
\begin{tabular}{ l l r r r l l r r r r l l }
\cline{1-13}
\noalign{\smallskip} 
PN\,G & Name & $\log\,T_{\rm b-b}$ & $\log {L} $ & Dist. & R$_{\rm out}$ &
M$_{\rm ion}$ & $V_{\rm av}$ & $V_{trb}$ & $M_{core}$ & $t_{dyn}$ & Central & Remarks \\
& & [K] &[L$_\odot$] & [kpc] & [pc] & [M$_{\odot}$] & \multicolumn{2}{c}{[km\,s$^{-1}$]} & [M$_{\odot}$] & [kyrs] & star & \\ 
\noalign{\smallskip}
\cline{1-13}
\noalign{\smallskip}
%%%%%%%%%%%%%%%%%%%%%%%   Teff   logL   dis   Rout   Mion    Va    Vt     Mcore    tdyn  cspn    %%%%%
000.9-04.8 & M 3-23    &  5.18 & 3.0 &  4.0 &  .11 &  .10 &  24  & 13   & 0.61   & 5.1 &       & \\
002.6-03.4 & M 1-37    &  4.40 & 3.9 &  8.0 &  .04 &  .08 &  27  &  0   & 0.60   & 1.7 &[WC11]?& \\
002.7-04.8 & M 1-42    &  4.92 & 3.0 &  4.0 &  .08 &  .16 &  13  &  0   & 0.60   & 5.4 &       & \\    
003.6+03.1 & M 2-14    &  4.64 & 3.0 &  8.0 &  .05 &  .06 &  17  & 10   & 0.60   & 2.9 & wels  & 6 lines \\
004.6+06.0 & H 1-24    &  4.56 & 3.8 &  7.0 &  .08 &  .13 &  24  &  0   & 0.58   & 3.7 & wels  & \\
006.4+02.0 & M 1-31    &  4.76 & 3.8 &  8.0 &  .06 &  .29 &  19  &  0   & 0.61   & 3.2 & wels  & \\
008.2+06.8 & He 2-260  &  4.80 & 3.2 & 12.0 &  .06 &  .13 &  17  &  0   & 0.61   & 3.5 &       &\\
211.2-03.5 & M 1-6     &  4.60 & 3.5 &  4.0 &  .03 &  .05 &  24  &  0   & 0.62   & 1.4 &       &\\
217.4+02.0 & St 3-1    &  4.96 & 2.5 &  4.0 &  .14 &  .20 &  26  &  0   & 0.60   & 6.1 &       &\\ 
232.0+05.7 & SaSt 2-3  &  4.80 & 2.5 &  4.0 &  .02 &  .01 &  19  &  0   & 0.64   & 1.1 &       &\\ 
232.4-01.8 & M 1-13    &  5.04 & 2.9 &  4.0 &  .10 &  .19 &  14  &  0   & 0.60   & 6.5 &       &\\
253.9+05.7 & M 3-6     &  4.72 & 4.0 &  1.5 &  .036&  .03 &  18  &  0   & 0.61   & 2.0 & wels  & \\
258.1-00.3 & He 2-9    &  4.73 & 2.9 &  1.5 &  .015&  .01 &  25  &  0   & 0.65   & 0.7 & wels  &\\
261.6+03.0 & He 2-15   &  5.17 & 3.0 &  2.2 &  .13 &  .32 &  22  &  0   & 0.71   & 6.4 &       &\\
283.3+03.9 & He 2-50   &  5.12 & 2.9 &  5.1 &  .14 &  .32 &  17  &  0   & 0.59   & 8.1 &       &\\
292.4+04.1 & PB 8      &  4.76 & 3.6 &  5.0 &  .06 &  .18 &  22  &  9   & 0.61   & 2.9 &[WC5-6]& \\
300.7-02.0 & He 2-86   &  4.83 & 3.4 &  3.0 &  .03 &  .05 &  14  &  11  & 0.62   & 2.0 &[WC4]  &\\
304.8+05.1 & He 2-88   &  4.72 & 2.8 &  4.0 &  .03 &  .03 &  23  &  0   & 0.62   & 1.4 &       & \\
309.0-04.2 & He 2-99   &  4.49 & 2.7 &  4.0 &  .16 &  .25 &  45  &  10  & 0.57   & 4.6 & [WC9] & \\
316.1+08.4 & He 2-108  &  4.51 & 3.8 &  4.0 &  .10 &  .14 &  21  &  9   & 0.57   & 5.0 & wels? & \\
321.0+03.9 & He 2-113  &  4.48 & 2.8 &  2.0 &  .01 &  .003&  18  &  15  & 0.64   & 0.6 & [WC10]& \\
325.0+03.2 & He 2-129  &  4.87 & 3.8 &  7.0 &  .03 &  .08 &  18  &  13  & 0.63   & 1.7 &       & \\
325.8+04.5 & He 2-128  &  4.70 & 3.3 &  5.0 &  .06 &  .13 &  12  &  10  & 0.59   & 4.3 &       & \\
347.4+05.8 & H 1-2     &  4.96 & 3.9 &  7.0 &  .02 &  .06 &  13  &  10  & 0.64   & 1.4 & wels  & 3 lines\\
351.1+04.8 & M 1-19    &  4.72 & 3.2 & 11.0 &  .10 &  .20 &  26  &  0   & 0.61   & 4.3 & wels  & 5 lines\\
352.1+05.1 & M 2-8     &  5.11 & 3.2 &  7.0 &  .06 &  .16 &  14  & 14   & 0.61   & 3.9 &[WO2-3]& 7 lines\\
355.9+03.6 & H 1-9     &  4.58 & 4.0 &  9.0 &  .04 &  .09 &  20  &  0   & 0.61   & 2.1 &       & \\
356.1+02.7 & Th 3-13   &  5.05 & 3.3 &  8.0 &  .03 &  .05 &  32  &  0   & 0.65   & 1.1 & wels  & 3 lines\\
356.5-02.3 & M 1-27    &  4.41 & 3.6 &  3.0 &  .04 &  .05 &  21  &  0   & 0.60   & 2.0 &[WC11]?& \\
357.1+03.6 & M 3-7     &  4.72 & 3.2 &  4.0 &  .05 &  .07 &  23  &  0   & 0.61   & 2.4 & wels  & 3 lines\\
357.4-03.2 & M 2-16    &  4.95 & 3.3 &  7.0 &  .09 &  .25 &  21  &  14  & 0.61   & 4.5 &       & \\
357.4-03.5 & M 2-18    &  4.64 & 3.5 &  7.0 &  .06 &  .09 &  24  &  0   & 0.60   & 2.8 &       &\\
359.1-02.3 & M 3-16    &  4.57 & 3.8 &  7.0 &  .08 &  .14 &  30  &  0   & 0.59   & 3.1 &       &\\
\noalign{\smallskip}
\cline{1-13}        
\end{tabular}       
\end{flushleft}     
\label{lista}
\end{table*}        

Figures \ref{li_1}--\ref{li_3} present the observed and modelled nebular
emission lines $\ion{H}{i}$ 6563\AA\ and [$\ion{N}{ii}$] 6583\AA. Although
for the indicated in the last column of the Table\,\ref{lista} six PNe
more lines were used for model fitting ([\ion{O}{iii}] 5007\AA\ and in some
cases [\ion{O}{i}] 6302\AA, [\ion{S}{ii}] 6732\AA, [\ion{S}{iii}] 6311\AA,
\ion{He}{ii} 4686\AA), we present only the two lines for a consistent
presentation. We do not draw the model radial distributions of the
velocity, density and surface brightness because our main results are the
mass-averaged expansion velocities.  The [WR]-subclass and {\it wels}
classification is adopted from Acker \& Neiner (2003) and for the lowest
temperature [WC11] stars from G\'orny et al. (2004). The summary of nebular and
central star parameters are listed in Table\,\ref{lista}. The last column
indicates whether more than two lines were used for modelling.  A few objects
are discussed in the following.

{\it PN\,G\,002.6-03.4 (M 1-37), [WC11]``?'')}. The density structure is
adopted to agree approximately with the image of Sahai (2000). The
brightest nebular structure is the inner ellipsoidal ring $1.5 \times
2.5\arcsec$ which is modelled by our $2\arcsec$ sphere. The multipolar
lobes as well as the extended spherical halo are beyond our analysis. We
obtained a lower temperature than Zhang \& Kwok (1991) for this low
excitation nebula. 

{\it PN\,G\,253.9+05.7 (M 3-6, wels)}. The radio image (Zijlstra et al. 1989)
reveals a complicated structure with four maxima.  Our spherical model is
density bounded, but the line ratios are not well fitted and suggest a mixture
of ionization and density boundaries. A small central cavity is indicated. The
best velocity field which reproduces the multicomponent structure of the
[\ion{N}{ii}] line shows multiple maxima.  Considering the complicated nebular
image our model should be treated with caution, but because the spectral
lines are almost symmetric the mass-averaged expansion velocity is believed to
be reliable.

{\it PN\,G\,321.0+03.9 (He 2-113, [WC 10])}. The HST image  (Sahai et al.
2000) reveals complex, highly aspherical structures which makes
spherical modelling doubtful. Nevertheless the average velocity still
is a useful parameter. In our observations this PN is unresolved and
the emission lines are symmetric.

\section{Derived parameters}
             
The newly analyzed data is combined with an earlier sample discussed in
Gesicki et al. (2003). Together with the 33 new PNe, the full sample contains
101 objects (from 73 earlier objects we removed three extragalactic PNe, and
two PNe are reanalyzed with the new spectra). The full sample contains 23
[WR], 21 {\it wels} and 57 non-emission-line central stars.

Several objects were reclassified based on the work of Acker \& Neiner
(2003) and G\'orny et al. (2004).  The distinction between emission-line
stars and non-emission-line stars depends on the depth of the available
spectra. It is therefore possible that some of the non-emission-line
stars would be classified as {\it wels} stars with deeper spectra. The
effect of detectability of the stellar lines on the classification is
discussed by G\'orny et al. (2004).

\subsection{Temperatures}

 \label{temps}

\begin{figure}
   \includegraphics[width=88mm]{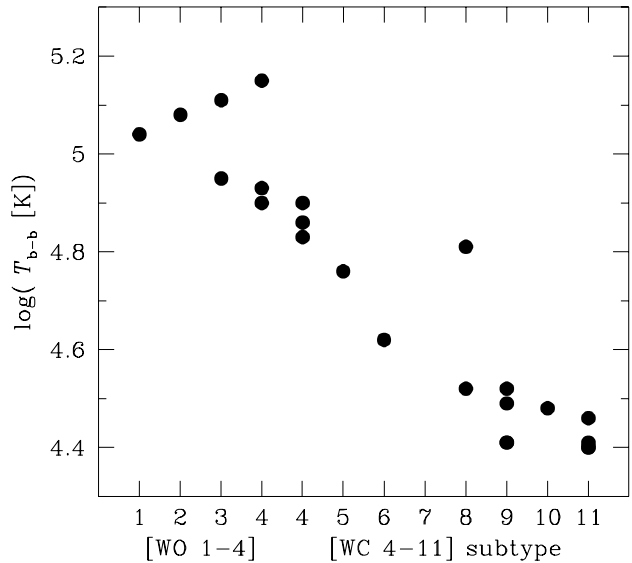}
      \caption{Stellar photo-ionization temperatures versus [WO]-[WC] 
               subclass. }
         \label{wct}
\end{figure}
   
The subclass of the [WR] stars provides an indication of stellar
temperature, but it can not be used to predict unique temperatures. This can
for instance be seen for the earlier subclasses (hotter stars) from the data
in Acker \& Neiner (2003). De Araujo et al. (2002) show that deep spectra
are needed for accurate association of a [WR] subclass. In this paper we use
the photoionization (equivalent black-body) temperature $T_{\rm b-b}$,
rather than the [WR] subclass. This parameter has the further advantage that
it is also available for the non-[WR] stars.  Fig.\,\ref{wct} shows the
relation between photoionization temperature and [WR] ([WO] and [WC])
subclass. There is a good relation overall, but the change in temperature is
smaller for the later subclasses (8--11). One discrepant object is M2-43
(27.6+04.2) where the temperature is much higher than the subclass ([WC
7-8]) would suggest.  The photo-ionization model fits the line ratios
reasonably well, and the strong [\ion{O}{iii}] line with weak
[\ion{N}{ii}] require a high temperature (Acker et al. 2002). Such high
temperature was also obtained by non-LTE model analysis of Leuenhagen \&
Hamann (1998). This object shows a dual dust composition (G\'orny et al.
2001). The [WC] subclass should be checked.

\subsection{Expansion velocities}

   \begin{figure*} 
%   \sidecaption
   \includegraphics[width=12cm]{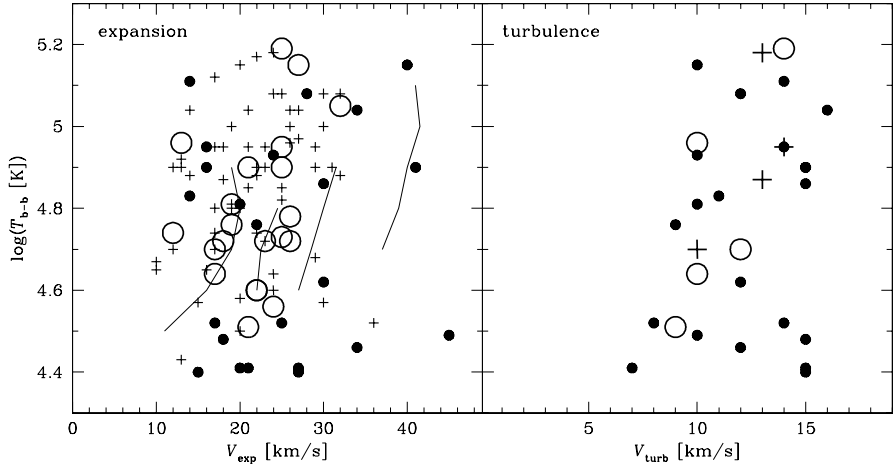} 
   \caption{The nebular
   expansion (left) and turbulent (right) velocities versus stellar
   temperature. The filled circles present [WR]-type PNe, open circles
   {\it wels} stars, and pluses show non-emission-line stars. In the
   left panel the  lines represent the sequences modelled by
   Sch\"onberner  et al. (2005a), each line for different initial density
   distribution. The accuracy of $V_{\rm av}$ is about
   $\pm 2{\rm km\,s^{-1}}$.} 
   \label{velocities} 
   \end{figure*}
   
For the whole sample of 101 PNe, the average (and the median) value of the
mass-averaged expansion velocities is $22\, {\rm km\,s^{-1}}$ with standard
error on the mean of $0.67 {\rm km\,s^{-1}}$. The average $V_{\rm av}$ for
[WR] PNe is $25\pm 2\, {\rm km\,s^{-1}}$ while for {\it wels} it is $22\pm
1\, {\rm km\,s^{-1}}$. The 1-$\sigma$ difference in expansion velocity is
much smaller than found in Pe\~na et al. (2003a), who find an average
expansion velocity for non-WR PNe of $21\,\rm km\,s^{-1}$ (virtually the
same as ours) but for [WR] PNe $36\,\rm km\,s^{-1}$. The difference can be
explained by their method of estimating the velocity: they use the
half-width at half maximum of the nebular lines and this is affected by the
turbulence in the velocity fields (Acker et al. 2002).

It is natural to explain the higher than average expansion velocity of
[WR] PNe in terms of stronger stellar wind acting on the swept-up shell.
Mellema \& Lundqvist (2002) studied such interactions and confirmed the above
tendency providing an example of [WR] PN expansion of $21\,\rm km\,s^{-1}$
versus $17\,\rm km\,s^{-1}$ for non-[WR] PN. This difference is smaller than
obtained by Pe\~na et al. (2003a), the values are rather comparable to ours.

Pe\~na et al. (2003b) argue that the expansion velocities for evolved [WR] PNe
are larger than for younger objects. In that paper the expansion velocity
is defined as half-width at 1/10 of the maximum intensity, a measure introduced
by Dopita et al. (1988). At this intensity level the emission lines are strongly
broadened by a turbulent component common for [WR] PNe so it is no surprise that
our values (corrected for the turbulence) are smaller. Pe\~na et al. (2003b)
arrive at their conclusion by plotting stellar temperatures versus expansion
velocities. In our sample, no such effect is visible (Fig.\,\ref{velocities},
left panel), except that the three highest expansion velocities in the sample
are from [WR] stars.  The $V_{\rm exp}$ and $T_{\rm b-b}$ values are spread over
the whole range. However, there is a trend for the [WR] stars to dominate
towards the lower-right corner in the figure, i.e. lower values of $T_{\rm
b-b}/V_{\rm exp}$. This trend looks very similar to that from Fig.\,3 of
Pe\~na et al. (2003b). However their [WR] PNe clustering in the upper-right
corner is not so similar.

Sch\"onberner et al. (2005a) discuss their observations of 13 PNe and
concluded that expansion speed increases with the $T_{\rm eff}$ of the
central star, i.e. with the nebular age. Such a situation cannot be
explained with simple models which assume an initial nebular density
distribution described by a power law $\rho \propto r^{-\alpha}$ with a
fixed index $\alpha$. Their conclusion is that the power-law index
increases systematically with $T_{\rm eff}$ (see their Fig.\,12). We plot
their model sequences for four values of the power law index in the left
panel of Fig.\,\ref{velocities}. The indexes are from left to right:
2.00, 2.5, 3.0 and 3.25. Our data for 101 PNe overlap with the lines (the
velocities are not defined exactly in the same way). We do not confirm the
Sch\"onberner et al. (2005a) conclusion, but the range of power law indices
which they derive provides a good fit to the spread seen in our sample.  The
trend in Fig.\,\ref{velocities} that [WR] stars dominate the distribution
towards lower values of $T_{\rm b-b}/V_{\rm exp}$, in combination with the
slope of the model tracks, suggests that some [WR] PNe may have steeper
initial density distributions than non-emission-line PNe.

\subsection{Dynamical ages}

The expansion velocity and radius of the nebula can be used to derive
the age of the nebula. One can use the outermost radius and the maximum
expansion velocity, but this is affected by the acceleration caused
by the overpressure in the ionized region. Instead we apply the
mass-averaged velocity, and use for the radius 0.8 of the outer radius,
roughly corresponding to the mass-averaged radius. To account for the
acceleration of the nebula we average between the current velocity and
that of the original outflow velocities on the Asymptotic Giant Branch.
The procedure is described in Gesicki et al. (2003).  We made
a simplifying assumption about the AGB outflow velocity setting it equal
to $10\,\rm km\,s^{-1}$ for all objects.

The dynamical ages can directly be compared between the different groups of
stars. However, they may not be equal to the true ages of the nebulae, due to
the various assumptions made. This problem was discussed together with
non-LTE analysis of some central stars of PNe. Rauch et al. (1994) obtained for
the object K\,1-27 that the dynamical age was more than one order of magnitude
smaller than the evolutionary age. In another paper Rauch et al. (1999) analyzed
four more PNe and obtained dynamical ages a few times larger than evolutionary
ages. The nebular and stellar ages were compared on a much bigger sample by
McCarthy et al. (1990). They obtained that generally the empirical ages were
larger than the evolutionary ones however their method for estimating dynamical
ages was recently strongly criticized by Sch\"onberner et al. (2005b). The
contemporary situation is far from being conclusive.

In the recent article we found a nice support for our method for estimating
nebular ages. Sch\"onberner et al. (2005b) show that both radii and velocities
increase gradually with time what justifies our averaging between original and
current velocity. For ionization bounded models (dominant in our sample) the
ages derived from ions [\ion{O}{iii}] and [\ion{N}{ii}] diverge in opposite
directions from true ages what supports our multi-line analysis where such
effects compensate. For density bounded PNe the situation is even better since
both lines result in the same age equal to the evolutionary one.

The dynamical age in combination with the stellar temperature gives the rate at
which the central star is increasing in temperature. This rate is very sensitive
to the core mass of the star, with more massive stars evolving much faster. To
assign a core mass, we interpolate between published tracks (see e.g G\'orny et
al. 1997, Frankowski 2003). For a given dynamical age and stellar
temperature we obtain the core mass (and luminosity). The dynamical ages and
core masses are listed in Table\,\ref{lista}. As for the ages, the mass
determinations are internally consistent between different objects, but may show
systematic offsets with respect to the true masses. As a check of the
consistency of the procedure the luminosities can be compared. In Gesicki et al.
(2003) we obtained that the luminosities from photoionization modelling are in
most cases smaller than those interpolated from evolutionary tracks. However this
can be interpreted as leaking of the nebulae: the PN does not intercept and
transform all stellar ionizing flux. This points to asymmetry or clumping.

\subsection{Turbulence}

For a number of nebulae, satisfactory fits to the velocity fields could not be
obtained.  In these cases we used enhanced turbulence, above what is expected
from thermal broadening and instrumental resolution.  Indicative of turbulence
is that lines of all ions show Gaussian shapes of comparable width:
thermal broadening gives much wider lines for the lightest element (hydrogen).
Comparing the H$\alpha$ with metal lines is here crucial because
differences in thermal broadening between different metal ions are small.
Therefore we searched for turbulent solutions only for those PNe with H$\alpha$
and at least one metal line available.  Turbulence was indicated for 30 of the
101 nebulae in the full sample.

Gesicki \& Acker (1996) and Acker et al. (2002) find that [WR]-type PNe
are characterized by strong turbulent motions while non-emission-line
PNe are not. This general tendency is confirmed in the present analysis,
but there are some exceptions. 

Among the {\it wels} we find both turbulent and non-turbulent objects.
Our full sample contains 21 {\it wels} of which 5 show turbulence and 16
do not. Of these 21 objects, 10 come from the new observations described
here (3 turbulent and 7 non-turbulent).  

We also find well-determined, turbulent solutions for 4 PNe not classified as
[WR] nor {\it wels} (M\,2-16, M\,3-23, He\,2-128, He\,2-129).  Not all
[WR]-type PNe in the new sample show turbulent motions: M\,1-27 and M\,1-37
have good solutions, without requiring turbulence.  The two non-turbulent [WR]
PNe are both classified by G\'orny et al. (2004) as '[WC11]?'.  They are the
coolest objects in our sample.  However, the strong correlation between
emission-line stars and turbulence remains: the fraction showing turbulence
for non-emission-line stars, {\it wels} and [WR] stars is 7\%, 24\%\ and 91\%,
respectively.

Uncertainties in the [WR] identifications may affect these fractions. The
turbulent non-WR PNe could have emission-line central stars.  This is
unlikely for M\,2-16, a well-studied, highly metal-rich PN (Cuisinier et al.
2000). He\,2-128 and He\,2-129 were studied spectroscopically by Dopita \&\
Hua (1997): the non-WR identification cannot be attributed to a lack of
observations, but faint stellar lines may have been missed.  Some objects
may have too few nebular lines to be able to detect turbulence: a velocity
gradient broadens a single line profile similar to turbulence -- we only
accepted turbulent solutions where they fitted clearly better the line
shapes than non-turbulent ones.  Deeper spectra would help here.

The onset of the turbulence and the [WR] phenomenon may not be completely
simultaneous. However, we find no evidence that the turbulence increases
for more evolved objects: Fig.\,\ref{velocities} (right panel) shows no
relation between turbulence and stellar temperature.

In trying to understand why the presence of turbulence in {\it wels} PNe is so
mixed, let us consider a possibility that a PN can switch between turbulent and
non-turbulent regimes during its lifetime. The transition time for this should
be comparable to the shock travel time across the nebula. Assuming a radius of
0.1\,pc and isothermal sound speed of $10\,{\rm km\,s^{-1}}$ we obtain a time of
$10^4$ years. Perinotto et al. (2004) stated that the shock front velocity can
be 3--4 times faster than the isothermal sound speed: this reduces the time
to a few thousands years. This is comparable to the ages of our PNe.  In our
sample we have a couple of (compact) objects with an age of about 1000 years
which have already developed turbulence. Acker et al. (2002) proposed that
turbulence in PNe is triggered or enhanced by [WR] stellar wind
inhomogeneities. The nebulae surrounding very active [WR] cores are
turbulent. The less active {\it wels} with much weaker winds, if they
could change their wind characteristics, they would develop turbulence, or
exhibit decaying turbulence, over the lifetime of the PN. Thus, the mixed
occurrence of turbulence in {\it wels} is compatible with a transitory
status, not unlikely a [WR]-star progenitor, provided the life time of the
phase does not exceed $\sim 10^3\,$yr.

\subsection{IRAS colours}

\begin{figure*}
%      \sidecaption
      \includegraphics[width=12cm]{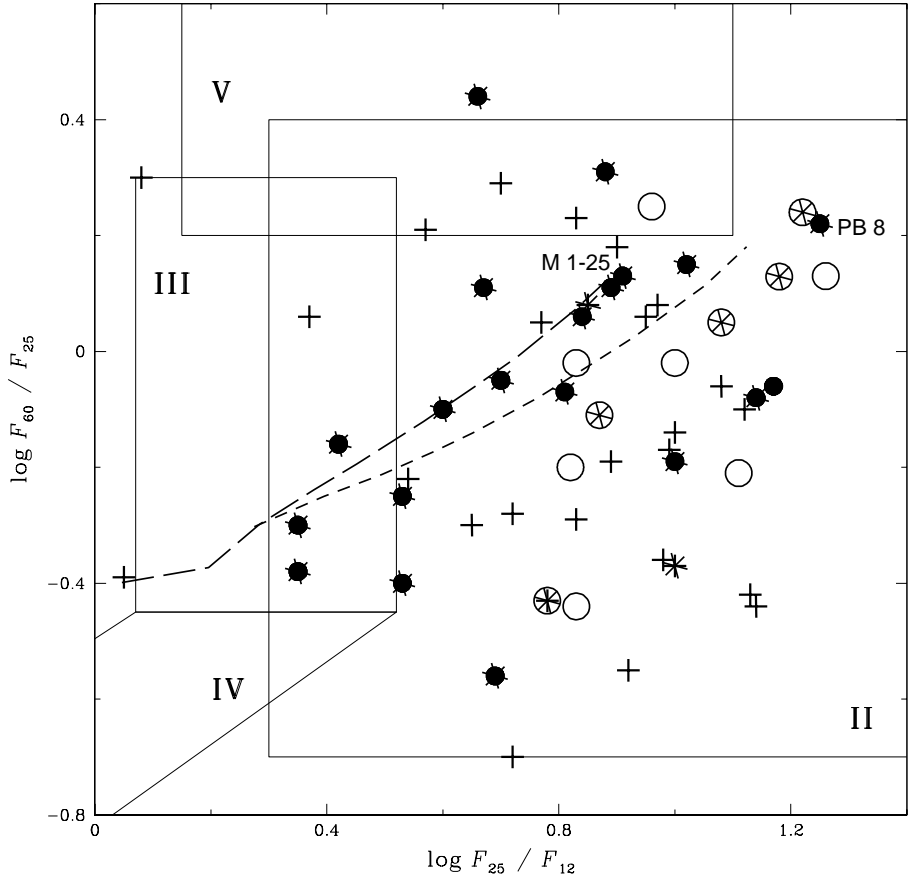}
      \caption{IRAS colours for the [WR] stars (filled circles, {\it
  wels} (open circles) and non-emission-line stars (plus symbols), with
  asterisks are overplotted the turbulent PNe. The boxes indicate source
  classifications  for different colours, taken from Zijlstra et al.
  2001. Box II are planetary nebulae,  box III post-AGB stars, box IV
  OH-Miras and Box V bipolar outflow sources.  The dashed lines show
  simple evolutionary sequences, explained in the text. Labeled are two
  individual PNe discussed in Sect.6.2.}
      \label{iras}
\end{figure*}
   
Planetary nebulae are strong infrared emitters, due to their heated
dust.  The dust colours follow the evolution: as the nebula expands,
the dust cools and the dust colours redden. Zijlstra (2001) has shown
that [WR] stars are stronger infrared emitters than other PNe, with a
tendency for bluer colours. The brightest infrared PNe are the IR-[WR]
stars, a subgroup of the cooler [WC] stars.

Fig.\,\ref{iras} shows the colour-colour diagram for the present sample.
Different symbols distinguish the [WR] stars (filled circles), the {\it wels}
(open circles) and the non-emission-line stars (pluses). In addition,
overplotted star symbols indicate turbulent nebulae.  We excluded the confused
source M\,2-6 where the IRAS colours appear to be those of a nearby AGB star.
The boxes are source classification regions of Zijlstra et al.  (2001).
Planetary nebulae fall in box II. The cooler [WC] stars are found mainly
towards the left in the diagram, in three cases with colours like those of
(young) post-AGB stars. (The three objects are BD +30\,3639, M\,4-18 and
He\,2-142.)  The {\it wels} are shifted to redder colours, and their
distribution is similar to those of normal PNe.  Note that the cooler
non-emission-line stars in our sample have 60/25-micron colours similar to the
cool [WC] PNe, but their 25/12-micron colours are very different.

We constructed dust models following the nebular expansion occurring as the
star heats up. We used a $r^{-2}$ wind, with the model and recipe of
Siebenmorgen et al. (1994). The model calculates the IRAS fluxes at any
particular time. The results are shown in Fig.\,\ref{iras}, for a carbon rich
(long dashes) and a oxygen rich (short dashes) nebular dust.  The evolutionary
progression from cool stars with bluer IRAS colours to hot stars with (more
evolved) redder IRAS colours is visible for the [WR] stars , but not for the
non-emission-line stars and the (less populated) {\it wels} group (Zijlstra
2001). 

Detection statistics show the differences between the groups. Out of the
101 stars, 60 have three good IRAS detections. The detection rates are
28 out of 57 non-emission-line stars, 12 out of 21 {\it wels} and 20 out
of 23 [WR] stars: respectively 49\%, 57\% and 87\%. This shows that the
[WR] stars are on average much stronger IRAS emitters. If we only
consider 25 and 60 micron, the detection rates become 89\%, 86\% and
100\%, i.e. the difference is rather reduced.  The numbers indicate that
[WR] stars are much brighter at 12-micron, but less so at the other bands.

The {\it wels} have the same detection rate as other non-[WR] stars.
However,  all 5 {\it wels} with turbulence have three-band detections;
their colours are similar to the [WO] stars. All {\it wels} without
12-micron detection are non-turbulent objects.

\section{Evolutionary parameters}

\subsection{Temperature distributions}

\begin{figure}
   \includegraphics[width=88mm]{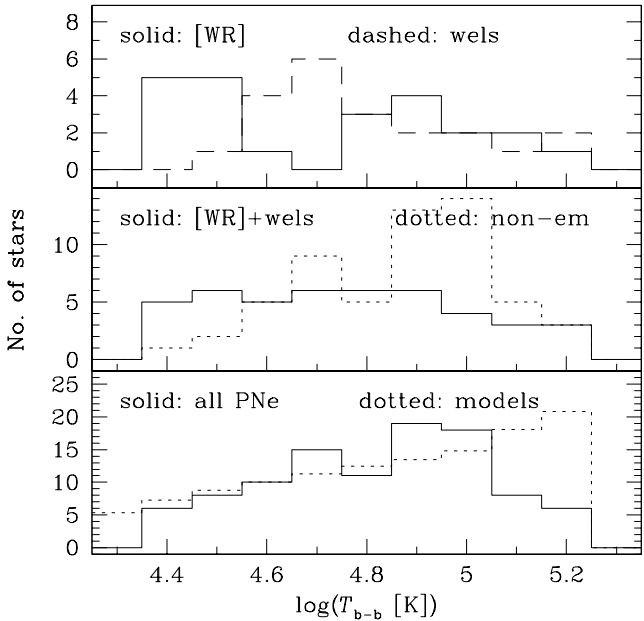}
   \caption{Temperature distribution of [WR], {\it wels} and
   non-emission-line stars, in logarithmic temperature bins. The bottom
   panel, discussed further in the text, compares the observed data with
   the histogram obtained from H-burning evolutionary models of
   Bl\"ocker (1995).} 
   \label{temp_lin} 
\end{figure}

The temperature distribution of the three groups of stars, defined as in section
\ref{temps}, is shown in Fig.\,\ref{temp_lin}.  The [WR] stars show some
clustering at low and high temperatures with a gap in between (see upper panel).
This gap has been noted before long time (see e.g. G\'orny 2001). The
{\it wels} stars, in contrast, peak precisely in this gap. In contrast, the
non-emission-line stars show a peak in their distribution at high temperature
(see central panel). 

Summing the two groups of emission-line stars (central panel), we find a flat
distribution (in logarithmic bins!).  Summing all three groups together
(bottom panel, drawn line) shows an increasing number of objects with $\log
T$.

We compare the temperature distribution of the various PNe samples with that
predicted by the evolutionary tracks of Blo\"ocker (1995). We interpolate
the tracks to constant temperature steps, in order to calculate the
predicted number of stars per temperature range (proportional to the time
spent in the range).  The model tracks have to first order a fairly constant
temperature increase, $dT/dt$. In logarithmic bins, this means that the
number of stars per bin increases with temperature. To be consistent
with our model analysis we only use the the horizontal part of the
evolutionary track, before the knee in the H-R diagram.
Fig.\,\ref{temp_lin} (bottom panel, dotted line) shows the predicted
distribution for hydrogen burning $0.605 {\rm M}_{\odot}$ track of Bl\"ocker
(1995).  The predicted number of objects (per logarithmic temperature bin)
gradually increases towards higher temperatures, as expected from the
increasing width of the bins and from slowing the evolution.

When compared to the observational data we see that no single group of PNe
produces a histogram similar to the models. The solid line in the bottom panel
of Fig.\,\ref{temp_lin} shows that the sum of all classes produce a histogram
most similar to the theoretically expected distribution.  Thus, the full
sample is representative for a uniformly sampled Bl\"ocker track. The flat
distribution of the emission-line stars implies that this group is biassed
towards lower temperatures, relative to standard evolution. The {\it wels} as
a separate group show a strong temperature preference and cannot represent a
full evolutionary track.

It is possible that our observations misrepresent the oldest PNe with the
hottest cores.  Our sample is mainly restricted to the spherical PNe with an
ionization boundary while for old PNe the ionization front breaks through and
may cause fragmentation of a nebula making it not suitable for our analysis.
We also cannot exclude a systematic error of our $T_{\rm b-b}$ values which
may shift the maximum of the observed distribution, and this error may also
depend on temperature.

\subsection{Dynamical ages of the PNe}

   \begin{figure*}
%      \sidecaption
   \includegraphics[width=12cm]{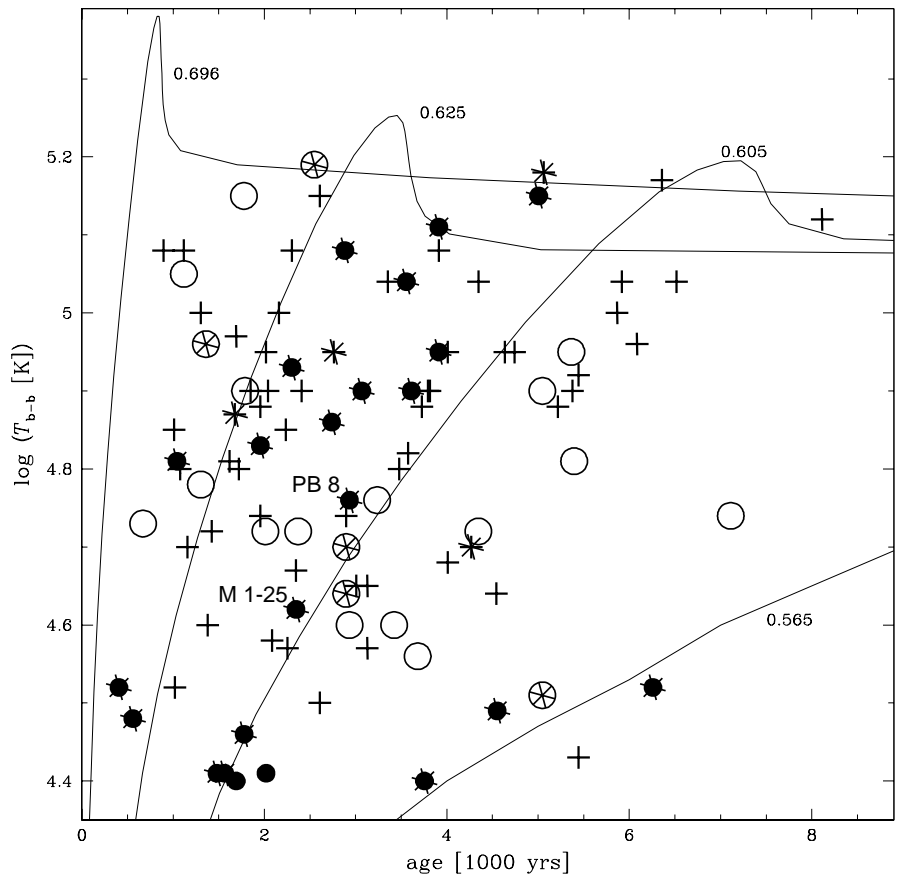}
      \caption{The age--temperature diagram. Filled circles indicate 
      [WR] stars, open circles are {\it wels} and pluses indicate 
      non-emission-line stars. Turbulent nebulae are indicated by
      the star symbols. Solid lines show Sch\"onberner's H-burning
      evolution. Labeled are two individual PNe discussed in Sect.6.2.
      }
         \label{evo}
   \end{figure*}

Evolutionary tracks should connect objects along sequences of increasing
dynamical age and increasing stellar temperature, until the stars reach the
knee in the HR diagram, after which the stellar temperature slowly
decreases.  Helium-burning tracks generally also follows this
sequence, but the stars evolve a few times slower (factor of 3 according to
Bl\"ocker 1995); the details depend on the precise timing of the thermal
pulse preceding the switch to helium burning.

Fig.\,\ref{evo} plots the dynamical age versus the temperature for the
stars in our sample.  The interpolated (hydrogen-burning) evolutionary
tracks are overplotted with a solid line. As in Fig.\,\ref{iras}
different symbols distinguish types of stars and indicate turbulence. 

The figure shows that some regions in this age-temperature plane are
dominated by certain types of objects. The [WR] stars are mainly found
in the narrow region  between the 0.605 and 0.625\,M$_\odot$ tracks,
with a few further objects at low temperatures and a range of ages. The
{\it wels} stars are spread mostly out at intermediate temperatures. The
non-emission-line stars tend to avoid the low temperature region, but
cover a region in the top-right corner which the other stars tend to avoid. 

The statistics of our full sample of 101 PNe shows that the average dynamical
age of non-emission-line central stars and of {\it wels} is 3200 years while
for [WR]-stars it is 2800 years, with standard error on the mean of about 300
years in both cases. This difference is not significant and the result may be
contaminated by a selection effect because it is easier to detect a [WR] star
at lower temperatures (see G\'orny et al. 2004).

\subsection{Non-emission-line objects}

In Fig.\,\ref{evo} the non-emission line objects are mixed with the
emission-line ones except of the upper right area. This is the place where
the cooling parts of evolutionary tracks of massive stars cross with heating
parts of evolutionary tracks of low mass stars. The hot ($10^5$\,K),
non-emission-line stars with old nebulae could be high-mass stars located on
the cooling track. The alternative interpretation is that they have
lower-mass central stars. In either case, the absence of emission-line stars
in this region is obvious. Either a minimum mass is required for a hot [WO]
star of about 0.61\,M$_\odot$, or the stellar wind ceases when the star
reaches the cooling track. The latter is plausible since the luminosity
drops quickly by a factor of 100 at this time. This fact is well known
for hot star winds (see e.g. Owocki 1994, Owocki \& Gayley 1995). Koesterke
\&\ Werner (1999) show that PG1159 stars have much weaker winds than {\it
wels} stars, and suggest that the wind mass-loss rate decreases rapidly when
the star enters the cooling track.

\subsection{Nebular evolution}

Sch\"onberner et al. (2005a) conclude that during the evolution the PN
material always enters the 'champagne phase' where the ionization front
breaks through and the PN becomes density bounded. This disagrees with
our result that most of the 101 PNe are ionization bounded despite 
their age and core $T_{\rm eff}$. In our sample the presence of the
ionization boundary was deduced from the strengths of the low-ionization
low-excitation lines. A possible explanation of this disagreement can be
found in Gesicki et al. (2003) where we compared the model luminosities
with those obtained from evolutionary tracks and suggested that 'PNe are
leaking'. This indicates that PNe are a mixture of regions transparent
and opaque for ionizing radiation. The spherical Sch\"onberner's and
Torun models cannot investigate asymmetric or clumpy nebulae. However
there is another explanation - the ionization boundary is not so
exceptional. The evolutionary sequence No.4 of Perinotto et al. (2004),
with high AGB mass loss rate and low AGB wind velocity, never becomes
optically thin and can serve here as a hint. It seems to be promising to
follow this idea and to search for other hydrodynamical models that
remain ionization bounded for a significant part of PN evolution.  In
Gesicki \& Zijlstra (2003) and Zijlstra et al. (2005 in preparation) we
presented models supported by HST images showing very small central
cavities of the nebulae. This finding supports the above idea of a low
AGB wind velocity, at least at the end of AGB and for some PNe.
Finding an appropriate hydrodynamic model might constrain the still
unknown AGB mass loss history.

\section{The hydrogen-poor sequences}

\subsection{The [WR] stars}

A global nebular quantity, turbulence, is predominant among the [WR] stars.
Mellema (2001) links the metals-enriched [WR] winds with long lasting
momentum-driven phase which drives the turbulence. This confirms that they
form a distinct group, and it can be interpreted as evidence for a separate
different evolutionary channel.  Fig.\,\ref{iras} and Fig.\,\ref{evo} show
that there is a marked uniformity among the majority of the [WR] stars.
Fig.\,\ref{evo} suggests this is related to the core mass of the star, where
stars of a relatively narrow range in core mass account for the majority of
the [WR] stars.  Caution should be taken when quoting precise masses: the
location in the age-temperature diagram depends on modeling assumptions as
discussed above.

The [WR] stars are in general hydrogen-poor, and are considered helium-burning
central stars. These stars will show a slower temperature increase than do
hydrogen-burning tracks of the same mass, although the detailed evolution is
complicated. The distribution of [WR] stars in Fig.\,\ref{evo} is not
inconsistent with such an inclined track, but strong conclusions are not
warranted.

There are three [WC]-PNe located near the 0.565\,M$_{\odot}$ track: from the
left M\,4-18, He\,2-99 and NGC\,40. All are old nebulae with cool stars.
The first is determined as old because of small expansion velocity, the
other two PNe are large ($\approx$0.15\,pc) as expected for old objects. Their
three central stars were analyzed by Leuenhagen et al. (1996). All have long
been suspected to be born-again PNe, where the central stars have suffered a
Very Late Thermal Pulse (VLTP), leading to a rejuvenated star inside an old
nebula (e.g.  Hajduk et al.  2005).  Helium-burning stars are not expected to
experience another VLTP. The stars along the VLTP sequence therefore represent a
separate evolutionary channel from the majority of the [WR] stars. The
difference may largely be one of timing, whether the last thermal pulse takes
place along the horizontal sequence (LTP), or on the cooling track (VLTP). The
only non-emission-line star in this region is the halo object PN\,G\,359.2-33.5,
which could well be a genuine low-mass object (McCarthy et al. 1991). The {\it
wels} in this region is He 2-108 (its classification is uncertain).

There is a group of 7 [WC] stars which combine very compact nebulae (low
ages) with very cool central stars. This group has temperatures in a narrow
range around $\log T_{\rm b-b} = 4.7$, well separated from hotter [WR]
objects (see Fig.\,\ref{wct} and Fig.\,\ref{evo}).  The IR-[WC] stars and
the mixed-chemistry objects are found here (Zijlstra 2001). It is not clear
that these belong to either of the two groups discussed above. The mixed
chemistry has been associated with binary evolution (Zijlstra et al. 2001,
de Marco et al. 2004), and in one case (CPD$-$56\,8032 --- Cohen et al.
1999, de Marco et al. 2002), there are strong indications for binarity. 
The precise role of a binary companion is not clear, but it may involve
retaining old ejecta in a disk, triggering accretion and possibly providing
a source of ionization. They could have similar origins to the R Cor Bor
stars (see the review of Clayton 1996), although the very different
circumstellar environments suggest the relation is not an evolutionary one.

This group of cool, compact objects could in principle have descendants
among the hot [WO] stars, which have more evolved nebulae. However, the
mixed chemistry (i.e. evidences for both a neutral oxygen-rich and an
inner carbon-rich regions) and unique IRAS colours argues against this. But
there are no obvious descendants among the non-emission-line stars either,
leaving the evolution of this group unexplained. If the stars are accreting,
the extended, cool photosphere could hide a much hotter underlying star:
this would explain the lack of intermediate-temperature related stars, but
still leaves the question of descendants open.

\subsection{The gap in [WR] temperature distribution}

The lack of PNe cores of [WC 5--7] type is well known (Crowther et al.
1998, G\'orny 2001). This feature can also be seen in our temperature
distributions presented in Fig.\,\ref{wct} and Fig.\,\ref{evo}. One of the two
[WR] star filling this gap is PB\,8 (PN\,G\,292.4+04.1) of [WC\,5-6] type (Acker
\& Neiner 2003). However by Parthasarathy et al. (1998) it was classified as
{\it wels} and Mendez (1991) found it as H-rich. In Fig.\,\ref{iras} this
object is positioned among {\it wels} stars. Therefore we propose that PB\,8 is
a {\it wels} and no longer fills the [WR] temperature gap. Another object, the
only known [WC 6] type - M\,1-25 (PN\,G\,004.9+04.9) fits the centre of the gap.
The Torun models resulted in a stellar temperature $\log T_{\rm b-b} = 4.6$
(Gesicki \& Acker 1996).  However much higher temperatures have been published
for the M\,1-25 central star, reaching $\log T_{\rm b-b} = 4.9$ (Samland et al.
1992) based on the helium ionization.  In Fig.\,\ref{iras} this object is placed
amongst other regular [WR] objects.  Therefore we propose that M\,1-25 is a
[WC\,6] object (we don't question the WR classification) but its temperature may
be above the discussed gap.

The uncertain status of the interlopers strengthens the case that the gap in the
temperature distribution of [WR] stars is real and quite prominent: the region
$4.55 < \log T_{\rm b-b} < 4.8$ appears devoid of [WR] stars. This supports the
proposed separate evolutionary sequences of cool and hot [WR] stars.
The same conclusion was reached by Crowther et al. (1998) who applied a
unified classification scheme to massive WR and low mass [WR] stars and
realized that the mentioned above gap is not visible for WR objects.

\subsection{The wels}

The {\it wels} appear to be a mixed zoo.  We tentatively identify three
groups. First, a group which shows similarities in distribution in
Fig.\,\ref{evo} to the [WR] stars. These include one object among the VLTP
sequence, and approximately 5 objects which form a low-temperature
extension to the hot [WO] stars and may be their progenitors. Second, around
6 {\it wels} are found along the high-mass track ($>0.625\,\rm M_\odot$).
These we suggest are objects where the high stellar luminosity drives a
stellar wind--they may be H-rich and not related to the [WR] stars. Finally,
a group along a low-mass track ($M\approx 0.6\,\rm M_\odot$).  These could
represent failed [WR] stars, where the luminosity is insufficient to drive a
full WR-type wind.

The {\it wels} show on average a shift in location in
Fig.\,\ref{velocities}: they fall in a region consistent with a lower power
index than do [WR] stars. The nebulae around [WR] stars tend to be more
centrally condensed than those of {\it wels} stars.  This points to a
different mass-loss history and is consistent with the suggestion that some
fraction, possibly the majority, of {\it wels} are not evolutionarily related
to the [WR] stars.

The small group which may be [WO] progenitors suggests a sequence {\it
wels}--[WO], reflecting increasing temperature. This could reflect a
bi-stability jump, where the strength of a stellar wind suddenly changes when a
change in ionization equilibrium makes different lines act in the absorption of
radiation. A similar situation is known among early-type supergiants (e.g. Vink
et al. 1999, Tinkler \&\ Lamers 2002). But the latter authors find that among
PNe cores, the wind suddenly {\it decreases} by a factor of up to 100, precisely
at the temperature where we find the {\it wels}-[WO] transition. They attribute
this to a CNO bi-stability jump. Their analysis used stars classified as O(H)
only, and could be biassed if hotter stars with high $\dot M$ appear as [WO]
stars.

For the other stars, the driving factor differentiating {\it wels} and [WR]
stars appears to be luminosity,

\subsection{The PG1159 stars}

The hot PG1159 stars are generally considered the most natural descendants of
[WR] stars (Werner 2001). They are not a uniform group: Werner et al. (1996)
refer to them as the 'PG1159 zoo'. There is also a group of so-called
'[WC]-PG1159 transition objects' which are found between [WCL] and [WCE]
groups in the $\log g-\log T_{\rm eff}$ plane (Werner 2001). Because our
results suggest separate sequences for cooler [WC] and hotter [WO], one could
envisage that cool [WC] stars evolve towards white dwarfs via a [WC]-PG1159
phase, whilst hot [WO] stars do so via a PG1159 phase. However, this remains
speculative.

It has been reported that the central star of the Longmore\,4 nebula changed
from a PG1159 type to a [WC 2-3], for a time of a few months (Werner et al.
1992). Its absorption spectrum transformed into an emission spectrum, with
changes proceeding on a timescale of days. It was interpreted as strongly
enhanced mass loss, and may have been triggered by an accretion event. As yet
it is the sole documented example of such impressive variability among PN
cores, but even so it shows that the picture is a complicated one, where the
chaotic picture of {\it wels} could be interpreted in terms of stellar wind
activity.

\section{Conclusions}

We have deduced for 33 PNe the velocity field. Since this was done in most
cases by considering two nebular lines we regard as robust the mass
averaged value only. For six PNe the analysis was based on three or
more lines and {\it is} therefore more reliable. The discussion of details of
velocity field like e.g. turbulence should be therefore treated with
caution while the analysis based on the mass averaged velocity (ages,
masses) should be reliable. We combined this sample with earlier
published data and discussed the full sample of 101 PNe.

We found no evidence for an increase in nebular expansion velocity with time.
This correlation was checked against $T_{\rm b-b}$ which should correspond to
stellar age and on the nebular dynamical age i.e. on two independent
characteristics. The [WR] stars show a tendency to low ratios of $T_{\rm
b-b}/V_{\rm exp}$. Comparison with the Sch\"onberner et al. (2005a) models
suggests that these objects have more centrally condensed nebulae, indicative of
increasing mass-loss rate with time on the AGB. Emission-line stars are not
found on the cooling track.

We find a correlation between IRAS 12-micron excess and the [WR] stars.  This
excess is not shown by the {\it wels}.  The cause of 12-micron excess is
typically a population of small grains. These may form in the wind, or form in
the collision between the stellar wind and the nebula. Without exception, the
{\it wels} do not show the 12-micron excess.

We confirm the stong correlation between turbulence and [WR] stars.
However, the correlation is not one to one: there are a few non-emission-line
stars with turbulence, and a few [WR] stars without.  The {\it wels}
appear intermediate in occurrence of turbulence.

The distribution in the age-temperature plane suggests that there are several
groups of [WR] stars and {\it wels}. One group shows evolved nebulae with cool
stars and likely originates from a Very Late Thermal Pulse. The hot [WO] stars
show a narrow distribution corresponding to a small range in core mass.  Some
{\it wels} fall on the same sequence but at lower temperature--they may be the
[WO] progenitors. Other {\it wels} are distributed on higher and on lower mass
tracks, which lack [WR] stars. Finally, a group of cool [WC] stars with
IR-bright compact nebulae have no counterpart among any other group of PNe.

The relation between the [WR] stars and the {\it wels} is therefore a
complicated one, with neither group being uniform. For the majority of the
{\it wels}, there is evidence against them being evolutionarily related to the
[WR] stars. There is a noticeable apparent difference in core mass and dust
properties and nebular turbulence.  A small group (5 objects) of {\it wels}
have characteristics consistent with [WO] star progenitors. Apart from this
{\it wels}--[WO] sequence, the {\it wels} contain stars at both the high
and low luminosity end of the distribution.  The former are likely
stars where the strong radiation field drives the wind, but are not
necessarily H-poor. The latter may be H-poor, failed [WR] stars where the
luminosity is too low to maintain a strong [WR] wind. For the [WO]
progenitors, the determining factor appears to be stellar temperature: the
particular ionization balance may reduce the efficiency of the
radiation-driven wind.

The determined discontinuity in temperatures of [WR]-type stars is in favor of
the suggested separate evolutionary channels of cool and hot [WR] objects.

\begin{acknowledgements}
This project was financially supported by the ``Polish State Committee for
Scientific Research'' through the grant No. 2.P03D.002.025, by the CNRS
through the LEA Astro-PF programme, and by a NATO collaborative program
grant No. PST.CLG.979726. ESO provided support via its scientific visitor
program. The NTT observations were associated with observing program
67.D-0527.
\end{acknowledgements}


\begin{thebibliography}{}

\bibitem[1996]{AGC1996}  Acker, A., G\'orny, S. K., Cuisinier, F., 1996, A\&A
305, 944

\bibitem[2002]{AGGD02} Acker, A., Gesicki, K., Grosdidier, Y., Durand, S., 2002,
A\&A 384, 620

\bibitem[2003]{AN2003} Acker, A., Neiner, C., 2003, A\&A 403, 659

\bibitem[1995]{B95} Bl\"ocker, T., 1995, A\&A 299, 755

\bibitem[1995]{Ch95}  Charbonneau, P., 1995, ApJS, 101, 309.

\bibitem[2002]{Ch02}  Charbonneau, P., 2002, An Introduction to Genetic
Algorithms for Numerical Optimization, NCAR Technical Note 450+IA (Boulder:
National Center for Atmospheric Research)

\bibitem[1996]{C1996} Clayton, G. C., 1996, PASP, 108, 225

\bibitem[1999]{Cea1999}  Cohen, M., Barlow, M. J., Sylvester, R. J., Liu, X.-W.,
Cox, P., Lim, T., Schmitt, B., Speck, A. K., 1999, ApJ, 513, 135

\bibitem[1998]{CMB1998} Crowther, P. A., De Marco, O., Barlow, M. J., 1998,
MNRAS 296, 367

\bibitem[2000]{CMK2000} Cuisinier, F., Maciel, W. J., K\"oppen, J., Acker, A.,
Stenholm, B., 2000, A\&A, 353, 543

\bibitem[2002]{Araujo02}
de Araujo, F. X., Marcolino, W. L. F., Pereira, C. B., Cuisinier, F., 2002,
AJ, 124, 464

\bibitem[2002]{dMBC} de Marco, O., Barlow, M. J., Cohen, M., 2002, ApJ, 574, 83

\bibitem[2004]{deMarco2004} de Marco, O., Barlow, M. J., Cohen, M., Bond, H. E.,
Harmer, D., Jones, A. F., 2004, in: Asymmetrical Planetary Nebulae III: Winds,
Structure and the Thunderbird,  Ed. by M. Meixner, J.H. Kastner, B. Balick and
N. Soker.  ASP Conference Proceedings (San Francisco: Astronomical Society of
the Pacific)  Vol. 313, p.100

\bibitem[1997]{DH97} Dopita, M. A., Hua, C. T., 1997, ApJS, 108, 515

\bibitem[1988]{DMWF88}  Dopita, M. A., Meatheringham, S. J., Webster, B. L.,
Ford, H. C., 1988, ApJ, 327, 639

\bibitem[1996]{Dreizler1996} Dreizler, S., Werner, K., Heber, U., Engels, D.,
1996, A\&A, 309, 820

\bibitem[2003]{Frankowski2003} Frankowski, A., 2003, A\&A 406, 265 

\bibitem[1996]{GA96} Gesicki, K., Acker, A., 1996, Ap\&SS 238, 101

\bibitem[2003]{GZ00} Gesicki, K., Zijlstra, A. A., 2000, A\&A 358, 1058

\bibitem[2003]{GZ03} Gesicki, K., Zijlstra, A. A., 2003, MNRAS 338, 347

\bibitem[2003]{GAS96} Gesicki, K., Acker, A., Szczerba, R., 1996, A\&A 309,
907

\bibitem[2003]{GAZ03} Gesicki, K., Acker, A., Zijlstra, A. A., 2003, A\&A 400,
957

\bibitem[2001]{G2001} G\'orny, S. K., 2001, Ap\&SS, 275, 67

\bibitem[1997]{GST1997} G\'orny, S. K., Stasinska, G., Tylenda, R., 1997, A\&A
318, 256

\bibitem[2001]{GSST2001} G\'orny, S. K., Stasinska, G., Szczerba, R., Tylenda,
R., 2001, A\&A 377, 1007

\bibitem[2004]{GSE2004} G\'orny, S. K., Stasinska, G., Escudero, A. V., Costa,
R. D. D., 2004, A\&A 427, 231

\bibitem[2005]{HZH2005} Hajduk, M., Zijlstra, A. A., Herwig, F., et al., 2005,
Science Vol. 308, p.231

\bibitem[2001]{H2001} Herwig, F., 2001, Ap\&SS 275, 15

\bibitem[2001]{K2001} Koesterke, L., 2001, Ap\&SS 275, 41

\bibitem[1997]{KH1997} Koesterke, L., Hamann, W.-R., 1997, A\&A 320, 91

\bibitem[1999]{KW1999} Koesterke, L., Werner, K, 1999, ApJ, 500, L55 

\bibitem[1992]{Ko92} Koza, J., 1992, Genetic Programming, Cambridge, MIT Press
1992.

\bibitem[1998]{} Leuenhagen, U., Hamann, W.-R., 1998, A\&A, 330, 265

\bibitem[1996]{} Leuenhagen, U., Hamann, W.-R., Jeffery, C. S., 1996, A\&A,
312, 167

\bibitem[2003]{MA2003} Marcolino, W. L. F., de Araujo, F. X., 2003, AJ 126, 887

\bibitem[1990]{MMMea1990} McCarthy, J. K., Mould, J. R., Mendez, R. H.,
Kudritzki, R. P., Husfeld, D., Herrero, A., Groth, H. G., 1990, ApJ, 351, 230

\bibitem[1991]{McCarthy1991} McCarthy, J. K., Rich, R. M., Becker, S. R.,
Butler, K., Husfeld, D., Groth, H. G., 1991, ApJ, 371, 380

\bibitem[2001]{M2001} Mellema, G., 2001, Ap\&SS, 275, 147

\bibitem[2002]{ML02} Mellema, G., Lundqvist, P., 2002, A\&A, 394, 901

\bibitem[1991]{M1991} Mendez, R. H., 1991, in: G.Michaud and A.Tutukov (eds.),
Evolution of Stars: The Photospheric Abundance Connection, IAU Symp. No. 145,
p.375

\bibitem[1994]{O1994} Owocki, S. P., 1994, in: Pulsation; rotation; and mass
loss in early-type stars, Edited by Luis A. Balona, Huib F. Henrichs, and Jean
Michel Contel. IAU Symposium no. 162; Kluwer Academic Publishers; Dordrecht,
p.475

\bibitem[1995]{OG1995} Owocki, S. P., Gayley, K. G., 1995, in: Wolf-Rayet stars:
binaries; colliding winds; evolution, Edited by Karel A. van der Hucht and
Peredur M. Williams. IAU Symposium no. 163; Kluwer Academic Publishers;
Dordrecht, p.138

\bibitem[1998]{PAS1998} Parthasarathy, M., Acker, A., Stenholm, B., 1998, A\&A
329, L9

\bibitem[2001]{PMS2001} Pe\~na, M., Medina, S., Stasinska, G., 2001, A\&A, 367,
983

\bibitem[2003]{PMS2003a} Pe\~na, M., Medina, S., Stasinska, G., 2003a,
Rev.Mex.A.A. 15, 38

\bibitem[2003]{PMS2003b} Pe\~na, M., Medina, S., Stasinska, G., 2003b,
Rev.Mex.A.A. 18, 84

\bibitem[2004]{PSSC2004} Perinotto, M., Sch\"onberner, D., Steffen, M.,
Calonaci, C., 2004, A\&A 414, 993

\bibitem[1994]{RKW1994} Rauch, T., K\"oppen, J., Werner, K., 1994, A\&A 286, 543

\bibitem[1999]{RKNW1999} Rauch, T., K\"oppen, J., Napiwotzki, R., Werner, K.,
1999, A\&A 347, 169

\bibitem[2000]{S2000} Sahai, R., 2000, ApJ 537, L43

\bibitem[2000]{snw2000} Sahai, R., Nyman, L.-\AA., Wootten, A., 2000, ApJ 543,
880

\bibitem[1992]{SKAS1992} Samland, M., K\"oppen, J., Acker, A., Stenholm, B.,
1992, A\&A 264, 184

\bibitem[2005]{sjspca2005a} Sch\"onberner, D., Jacob, R., Steffen, M.,
Perinotto, M., Corradi, R. L. M., Acker, A., 2005a, A\&A 431, 963

\bibitem[2005]{sjs2005b} Sch\"onberner, D., Jacob, R., Steffen, M., 2005b, A\&A
441, 573

\bibitem[1994]{sie1994} Siebenmorgen, R., Zijlstra, A. A., Kr\"gel, E., 1994,
MNRAS, 271, 449

\bibitem[2002]{Tinkler2002} Tinkler, C. M., Lamers, H. J. G. L. M., 2002,
 A\&A, 384, 987

\bibitem[1993]{t1993} Tylenda, R., Acker, A., Stenholm, B., 1993, A\&AS 102, 595

\bibitem[1999]{V1999} Vink, J. S., de Koter, A., Lamers, H. J. G. L. M., 1999,
A\&A 350, 181

\bibitem[1989]{W89} Weinberger, R., 1989, A\&AS, 78, 301

\bibitem[1992]{Werner1992} Werner, K., Hamann, W.-R., Heber, U., Napiwotzki, R.,
Rauch, T., Wessolowski, U., 1992, A\&A, 259, L69

\bibitem[1996]{WDHR1996} Werner, K., Dreizler, S., Heber, U., Rauch, T., 1996,
in:  C.S. Jeffery and U. Heber (Eds.), Hydrogen deficient stars, ASP Conference
Series, Vol. 96, p. 267

\bibitem[2001]{Werner2001} Werner, K., 2001, Ap\&SS, 275, 27

\bibitem[1991]{ZK1991} Zhang, C. Y., Kwok, S., 1991, A\&A 250, 179

\bibitem[1989]{ZPB1989} Zijlstra, A. A., Pottash, S. R., Bignell, C., 1989,
A\&ASS 79, 329

\bibitem[1994]{ZHC1994} Zijlstra, A. A.,  van Hoof, P., Chapman, J. M., Loup,
C., 1994,  A\&A, 290, 228

\bibitem[2001]{ZCL2001} Zijlstra, A. A., Chapman, J. M., te Lintel Hekkert, P.,
Likkel, L., Comeron, F., Norris, F. P., 2001, MNRAS, 322, 280

\bibitem[2001]{Zijlstra2001} Zijlstra, A. A., 2001, Ap\&SS, 275, 79

\end{thebibliography}
\end{document}